\documentclass[12pt]{article}
\usepackage{amssymb,amsmath,amsfonts,amssymb}
\usepackage{cite}

\newlength{\dinwidth}
\newlength{\dinmargin}
\def\RR{{\mathbb R}}

\def\bx{{\boldsymbol x}}
\def\by{{\boldsymbol y}}
\def\bphi{{\boldsymbol \phi}}
\def\bpi{{\boldsymbol \pi}}
\def\bsigma{{\boldsymbol \sigma}}
\def\bdelta{{\boldsymbol \delta}}
\def\bzero{{\boldsymbol 0}}

\newcommand{\w}[1]{{\langle \, #1 \, \rangle }}

\def\ie{\textit{i.e.\ }}
\def\etc{\textit{etc}}
\def\eg{\textit{e.g.\ }}

\begin{document}
\title{Unruh Effect and the Concept of Temperature}
\author{Detlev~Buchholz \ and \ Christoph~Solveen \\
Institut f\"ur Theoretische Physik,
Universit\"at G\"ottingen,  \\ 37077 G\"ottingen, Germany}
\date{}
\maketitle
\begin{abstract} \noindent
Based on a discussion of the concepts of temperature,
passivity and efficiency in the framework of quantum field theory, the 
physical interpretation of the Unruh effect is reviewed. \\[2mm]
\end{abstract}
%
\noindent The Unruh effect~\cite{Un}, \ie the 
theoretical assertion that any uniformly accelerated observer
in an ambient (inertial) ground state 
is led to describe his measuring results 
in terms of a Gibbs ensemble,  
has been extensively discussed in the
literature, cf.\ the comprehensive reviews~\cite{Ta,CrHiMa} 
and references quoted there. 
Yet in spite of the fact that almost all
of its computational and formal
aspects have been considered, it seems that 
no consensus has been reached about its physical interpretation,
cf.~for example \cite{BeFeKaMuNa, FoOC, Ea}.
Does the Unruh effect mean that a thermometer waived around 
by an accelerated observer in empty space will
record a non--zero temperature? And, if not, then what is its physical 
significance?  The ongoing 
debate originates primarily from the 
frequently ambiguous usage of thermodynamic concepts
(such as temperature) in quantum field theory. It is the 
aim of this article to 
outline how these concepts are to be incorporated 
into the microscopic framework and to 
review the Unruh effect on this basis. We will describe 
in meaningful physical terms what the theory actually predicts.

It is appropriate to recall in this context the 
well known fact that there are two different  
phenomenological concepts 
of temperature \cite{Pl}. The first one enters in the zeroth law: 
Stationary states of physical systems can be subdivided into
classes whose members do not change if brought into thermal contact,   
they can coexist.
This observation is the basis for inventing thermometers, \ie
observables whose measured values in a given stationary state 
allow to determine the particular class to which it belongs. 
The measured values are then  taken as empirical temperature
of the state (simply called temperature in the following)
and it is this quantity which comes closest to the 
intuitive idea of temperature. The other, less intuitive 
notion is based on the second law. According to 
it it is impossible to gain energy by a cyclic 
operation which utilizes solely the thermal contact with a 
single equilibrium (hence stationary) 
state. Equilibrium states are therefore 
referred to as passive. Moreover, the maximum possible 
efficiency of cyclic operations which utilize 
the thermal contact with two equilibrium states 
is of a universal (system independent) nature and can 
in principle be determined experimentally. It allows to 
characterize each class of equilibrium states by another, more 
intrinsic parameter. For the sake of conceptual clarity 
let us call it Carnot--parameter. 
Under standard conditions this parameter can be determined 
by a Carnot process which, by its very 
definition, uses only thermal contacts and adiabatic changes 
of states, there may be no external forces. One can then establish a 
one--to--one correspondence between the 
temperature and the Carnot--parameter and rename
the latter ``absolute temperature''. This 
correspondence is lost, however, in the presence of 
external forces which affect the efficiency of cyclic 
processes and thereby the value of the Carnot-parameter. 
It is apparent that the latter point  matters in 
case of the Unruh effect. 

Next, let us outline in appropriate generality 
how the underlying phenomenological concepts are 
implemented in the theoretical setting of quantum field theory.
Since we will consider states of a given physical system at
different temperatures, requiring different Hilbert
space representations for their description, the algebraic 
approach to quantum field theory is appropriate here \cite{Ha}. 
There the (localized) measurements and operations which one can perform are 
described by operators which are elements of some 
specific  Algebra ${\cal A}$. We make
use here of the Heisenberg picture, where the states of the system,
\ie the (positive, linear and normalized)
expectation functionals $\w{\cdot}$ on ${\cal A}$,  are
time independent and the observables depend on time. 
Their evolution is given by the solutions of the Heisenberg equation, 
\begin{equation*}
{\textstyle \frac{d}{dt}} A(t) = \delta(A(t))  \, ,
\quad A(0) = A \in {\cal A} \, , 
\end{equation*} 
where the algebraic generator
$\delta$ depends on the microscopic interactions and 
on the motion of the observer\footnote{In a chosen Hilbert space of 
states the action of $-i \delta$ can often be described 
by a commutator with a suitable operator (Hamiltonian, Liouvillian \etc).
The result of this commutation does not depend on the particular  choice of 
a state space, however, and defines the algebraic generator $\delta$.}
and $t$ denotes his proper time. Thus we will have
to consider different generators, but may restrict 
attention to those cases where the generator does not depend 
explicitly on time.

By comparing expectation values at different times, an  
observer can identify the stationary states. They satisfy 
$\w{A(t)} = \w{A}$ for $A \in {\cal A}$, $t \in \RR$ or, 
equivalently, 
\begin{equation} \label{stationarity}
\w{\delta(A)} = 0  \quad \mbox{for} \  A \in {\cal A} \, .
\end{equation} 
Without changing mean values, 
the fluctuations of the observables $A$ 
can be suppressed  in these states  by taking time averages, 
$\bar{A}^{\, t}  \doteq 
{\textstyle \frac{1}{t}} \int_0^t ds \, 
A(s)$.
A stationary state is said to be extremal if these fluctuations completely 
disappear in the limit of large $t$, \ie 
$\lim_{\, t \rightarrow \infty} (\w{\bar{A}^{\, t} \bar{A}^{\, t}} 
- \w{\bar{A}^{\, t}}{}^2 \, ) = 0$. 
Such a state cannot be decomposed into a mixture of  
other stationary states. It is of interest in this context
that, quite generally, any stationary state is extremal or 
a mixture of extremal stationary states. 
Let $f$ be the number of observables 
$A_1, \dots , A_f \in {\cal A}$ needed to discriminate 
the extremal stationary states. Given such a collection of observables 
one could use the arrays of mean values 
\mbox{$(\w{A_1}, \dots , \w{A_f}) \in \RR^f$},
describing certain intensive properties of the states,
as their labels. We use here the shorthand notation 
$\w{\cdot}_\sigma , \  \sigma \in \Sigma^f$, where 
$\Sigma^f$ is any suitable index set.

The zeroth law corresponds to the observation
that there exist classes of extremal stationary states 
whose members can coexist if brought into thermal contact.
They correspond to subsets $\Sigma_\tau \subset \Sigma^f$
labeled by some real parameter $\tau$, the (empirical)
temperature. We do not need to discuss here, how these classes
are determined in a given model. But it is clear from the  
preceding discussion that the observables 
$A_1, \dots , A_f$ would suffice in order to decide to which 
class $\Sigma_\tau$ a given state $\w{\cdot}_\sigma$ belongs,
\ie to determine its temperature. In order to
simplify the discussion, let us assume that there 
exists  a single observable $\theta \in {\cal A}$
which likewise allows to determine the class of a stationary 
state. Thus its mean values coincide in all states of a given class, 
\begin{equation} \label{classes}
\w{\theta}_\sigma = \Theta_\tau \, , \quad \sigma \in
\Sigma_\tau \, ,
\end{equation}  
but they differ in states 
belonging to different classes. 
Any such observable corresponds to a thermometer and
its readings $\Theta_{\tau}$ can be interpreted as 
temperature on a  $\theta$--dependent temperature scale.

Not all extremal stationary states are 
equilibrium states in the sense of the second law. 
In order to identify the subset of equilibrium (passive) 
states in the theoretical setting one has to study the 
effects of cyclic operations on the states.
As has been shown by Pusz and Woronowicz \cite{PuWo}, 
the effect of such an operation after one (or several)
cycles can be described by the adjoint action of some unitary operator 
$U \in {\cal A}$. A given state $\w{\cdot}_\sigma$
is transformed by this 
operation into the state $\w{U^* \cdot U}_\sigma$
and the difference between the energy of the state 
after and before this operation is given by
\begin{equation} \label{passivity}
\Delta E(U)_\sigma = -i \w{U^* \delta(U)}_\sigma \, ,
\quad U = {U^*}^{-1} \in {\cal A} \, .
\end{equation} 
Since one cannot extract energy from equilibrium states   
by cyclic operations, these states can be 
identified in the theoretical framework by the passivity condition
$\Delta E(U)_\sigma \geq 0$ for all unitary operators $U \in {\cal A}$.
For given temperature $\tau$ the corresponding passive states
correspond to some subset $\Pi_{\tau} \subseteq \Sigma_\tau$.
According 
to the deep results of Pusz and Woronowicz, any such passive state
is either a ground state for the underlying dynamics, 
or a KMS state (which is the appropriate generalization of 
the notion of Gibbs state for systems in infinite volume \cite{BrRo}).  

The Carnot--parameter, which labels equilibrium states,
can be determined in the theoretical setting in several
ways. A physically transparent method, established by 
Sewell \cite{Se1}, cf. also~\cite{BrRo},
is based on the test of correlation inequalities. There one 
compares for given passive state $\w{\cdot}_\sigma$,
$\sigma \in   \Pi_{\tau}$ and 
any operator $A \in {\cal A}$ which is normalized 
according to $\w{A^*A}_\sigma = 1$  
the expectation values $\w{A A^*}_\sigma$ and 
$i \w{A^* \delta(A)}_\sigma$. The Carnot--parameter 
$C_\sigma$ of state $\w{\cdot}_\sigma$,  $\sigma \in   \Pi_{\tau}$ 
is the unique  energy value fixed by the inequalities  
\begin{equation} \label{carnot} 
C_\sigma \, \ln{\w{A A^*}_\sigma} \geq i \w{A^* \delta(A)}_\sigma 
\quad \mbox{for all normalized} \ A \in {\cal A} \, ,
\end{equation}
respectively $C_\sigma = 0$ if $\w{A^* A}_\sigma = 0$ for some $A \neq 0$.
Thus the Carnot parameter describes a global feature of the
respective state. Passive states having different Carnot--parameters 
in a given frame of reference cannot
coexist, \ie they have different temperatures. 

For passive states in an inertial frame, where the temperature 
is everywhere the same within the system, there is a one--to--one
correspondence between the Carnot--parameter and the temperature. 
It is then meaningful to take the  Carnot--parameter (which does not 
depend on the largely arbitrary choice of some thermometer observable) 
as absolute temperature. But this correspondence is lost for
passive states in non--inertial frames, where the
temperature may vary within the system. We will illustrate this
fact below. 

To summarize: Given the dynamics of any observer, described by 
some generator~$\delta$ acting on the algebra ${\cal A}$, 
one can~(1) identify states which are stationary, 
(2)~exhibit observables which allow to 
determine the temperature of states, 
(3)~identify equilibrium
states by a condition of passivity and (4)~determine 
the Carnot--parameters of equilibrium states
which can be interpreted as absolute temperature in the
absence of external forces. Since the corresponding 
mathematical conditions involve only 
local operators and the infinitesimal generator 
$\delta$ of  time evolution, the theory is consistent 
with the empirical fact that thermal 
properties of stationary states can be probed locally and 
at any given instant of time (replacing time averages
by ensemble averages). 

Having explained, how the thermodynamical concepts fit into
the framework of quantum field theory, let us consider now a 
concrete example. For simplicity, we take the model of a massless 
scalar free field in four--dimensional Minkowski space ${\cal  M}$. 
We use the metric $(+,-,-,-)$ and standard units $\hbar = c =1$ 
as well as $k=1$ (Boltzmann constant). Choosing proper 
coordinates $x = (x_0, {\bx})$ in some given Lorentz 
frame, the field ${\bphi}$ and its conjugate ${\bpi}$
at time $x_0 = 0$ satisfy the standard canonical 
commutation relations 
\begin{equation*} 
[\bphi(\bx) , \bpi(\by)] = i \, \bdelta(\bx - \by) \cdot  1 \, , \quad 
[\bphi(\bx) , \bphi(\by)] = [\bpi(\bx) , \bpi(\by)] = 0 \, . 
\end{equation*} 
After smearing with test functions, their products and sums  
generate a (kinematical) algebra 
${\boldsymbol {\cal A}}$, whose elements describe localized measurements 
and operations.

\noindent (I) \ Let us consider first the dynamics of the field, 
as seen by some inertial observer
in the given frame. Its generator $\delta_0$ is fixed by the
relations 
\begin{equation*} 
\delta_0(\bphi(\bx)) \doteq \bpi(\bx) \, , \quad 
\delta_0(\bpi(\bx)) \doteq  \Delta_\bx \, \bphi(\bx) \, ,
\end{equation*} 
 where $\Delta_\bx$ denotes the Laplacian. 
The well--known solution of the corresponding Heisenberg equations determine 
the field $\phi_0(t,\bx)$ at arbitrary spacetime points 
$(t,\bx) \in {\cal M}$, \ $t$ being the proper time of
the observer. It is given by 
\begin{equation*} 
\phi_0(t,\bx)   =
i \int \! d\by \, \dot{C} (t,\bx - \by) \, \bphi (\by)  +
i \int \! d\by \, C (t,\bx - \by) \, \bpi (\by) \, ,
\end{equation*} 
where $C$ is the commutator function  
\begin{equation*} 
C(x) \doteq  (2 \pi)^{-3} \int \! dp 
\, \varepsilon (p_{0}) \, \delta (p^{2}) \, e^{-i xp} \ = \ 
- i (2  \pi)^{-1} \, \varepsilon (x_0) \delta (x_0{}^2 - \bx^2) \, .
\end{equation*} 
There exist many extremal stationary states in this model. We restrict
attention here to 
the physically significant family of 
quasifree states which are fixed 
by the two--point functions\footnote{\ 
The reader will recognize that these states 
are commonly interpreted as equilibrium states of  
inertial observers. We want to outline here, how this
interpretation can be justified on the basis of the 
preceding discussion.}
\begin{equation*} 
\w{\phi_0(x) \, \phi_0(y)}_\sigma \doteq 
(2 \pi)^{-3} \int \! dp \,
\varepsilon(p_{0}) \, \delta (p^{2}) \,
(1-e^{- \sigma p})^{-1} \, e^{-i(x - y)p} \, , \quad \sigma \in
\Sigma^4 \, , 
\end{equation*}
where $\Sigma^4 \doteq \{ \, \sigma \in \RR^4 \, | \, \sigma_0 >
|\bsigma| \, \} $. Since these states satisfy the Hadamard
condition \cite{Wa}, it is meaningful to proceed from
${\boldsymbol {\cal A}}$ to a larger algebra 
${{\cal A}}$ containing also composite fields such 
as the stress energy tensor. In order to distinguish 
the stationary states one may use the first 
four components of the tensor field
\begin{equation*} 
\epsilon_{\mu \nu}(x) \doteq - (1/4) \ \partial_\mu \partial_\nu \, 
\big(\phi_0(x+z) \phi_0(x-z)
- h(x+z,x-z) \, 1 \big) \, |_{z = 0} \, ,
\end{equation*}
where $\partial$ denotes the gradient with 
respect to $z$ and
$h$ is the Hadamard parametrix 
for the wave equation \cite{Wa}; 
it compensates the leading
singularity of the product of fields in the indicated coincidence 
limit. In the present example this regularization 
amounts to the familiar Wick ordering on Fock space. The expectation values
of this tensor field resulting from the above two--point functions are  
\begin{equation*}
\w{\epsilon_{\mu \nu}(x)}_\sigma = 
(\pi^2/90) 
\big( 4 \sigma^\mu \sigma^\nu - \sigma^2 g^{\mu\nu} \big)
(\sigma^2)^{-3} \, , \quad \sigma \in \Sigma^4 \, .
\end{equation*}
In the Lorentz frame with time direction given by 
$\sigma$, this tensor is diagonal and depends only on $\sigma^2$.
Thus given $\sigma_1, \sigma_2$
with $\sigma_1^2 = \sigma_2^2$ the corresponding states differ 
only by their macroscopic flow velocities which, for an observer 
in the respective center of
mass frame, are equal and have opposite direction.
Since parity is an unbroken symmetry in the present model 
a space reflection has no effect on the thermal properties of 
states, so the two states can coexist in the sense of 
the zeroth law, \ie they have the same temperature.  The scalar field
\begin{equation*}
\theta_0(x) \doteq 
\big(\phi_0(x+z) \phi_0(x-z)
- h(x+z,x-z) \, 1 \big) \, |_{z = 0}  
\end{equation*} 
has the expectation values $\w{\theta_0(x)}_\sigma = (12 \,
\sigma^2)^{-1}$. 
Hence it may be used as a thermometer observable, determining the classes 
of coexisting states 
$\Sigma_\tau$, $\tau \geq 0$ by the condition (at any point $x \in {\cal M}$) 
\begin{equation*}
\w{\theta_0(x)}_\sigma = \tau \, , \quad \sigma \in 
\Sigma_\tau \subset \Sigma^4 \, . 
\end{equation*} 
Since for large $\sigma^2$ the corresponding functionals 
$\w{\cdot}_\sigma$ converge to the vacuum state, 
fixed by the two--point function 
\begin{equation*}
\w{\phi_0(x) \phi_0(y)}_\infty = (2 \pi)^{-3} \int \! dp \,
\theta(p_{0}) \, \delta (p^{2}) \, e^{-i(x - y)p} \, , 
\end{equation*} 
small values of  $\tau$ correspond to the idea of ``cold'' and large ones 
signify ``hot''.  

Next, one has to identify the subset 
$\Pi_{\tau} \subset \Sigma_\tau$ corresponding to the 
passive states for given temperature $\tau$. Here the time direction 
$e_0 = (1, \bzero)$ of the observer matters.
We omit the computations relying on the KMS condition 
\cite{BrRo} and only state the result:
For the given generator  $\delta_0$, 
the condition of passivity is satisfied by a state
$\w{\cdot}_\sigma$, $\sigma \in \Sigma^4$ if and only if 
$\sigma$ and $e_0$ are parallel. 
Hence $\Pi_\tau = \{ \mbox{\footnotesize $\sqrt{1/12\tau}$} \, e_0 \}$
consists of a single point, \ie 
there exists only one passive state for each temperature
in this model (which does not describe chemical potentials, 
different phases \etc). It is also evident on physical grounds that
all other states in  $\Sigma_\tau$ are not passive since, 
making use of their non--zero macroscopic flow velocities, the observer
could gain energy from these states by cyclic operations, think
of a pinwheel. 

It remains to determine the Carnot--parameter for the single
state associated with $\Pi_\tau$. The  result is  
$C_\tau = \mbox{\footnotesize  $\sqrt{12\tau}$}$ \ which, once again, 
can be derived from the KMS--condition \cite{BrRo}. Since there 
are no external forces, one can interpret this parameter as 
absolute temperature $T$, thereby establishing the relation 
$T = \sqrt{12\tau}$ between the 
temperature scale, based on the zeroth law
and our choice of a thermometer observable, and 
the absolute temperature scale, based on the second law. 

Of course, these results are well known in one way 
or another. The main point, illustrated by this example, 
is the insight that the 
physical interpretation of states can be deduced from
the theory in a systematic, unbiased manner. We will now 
discuss the Unruh effect from this perspective.

\noindent (II) \ We consider an observer who is at rest 
at the point $(0,g^{-1},0,0) \in {\cal M}$, experiences 
a constant acceleration $g>0$ into the $1$--direction and drags 
along his experimental devices on neighboring 
orbits.\footnote{Clearly, the observer could 
depart from any point of the $x_0 = 0$ plane with non--zero
velocity and be accelerated into any direction. This particular choice 
of initial conditions simplifies notation. The motion of 
the observer and of his experimental devices is adequately described 
by the Fermi--Walker transport which, from the inertial point of view,
is given by Lorentz boosts.} 
These observables are described by elements of the subalgebra 
${\boldsymbol {\cal A}_{\, g}} \subset {\boldsymbol {\cal A}}$
which is generated by the field and its conjugate smeared 
with test functions which have support in the region $x_1 > 0$ of
the $x_0 = 0$ plane. For the accelerated observer
the (kinematical)
observables maintain their inertial interpretation,
but their dynamics changes. It is given by the generator $\delta_g$ 
\begin{equation*}
\delta_g(\bphi(\bx)) \doteq g x_1 \bpi(\bx) \, , \quad 
\delta_g(\bpi(\bx)) \doteq  g x_1 \Delta_\bx \bphi(\bx) + g \partial_1
\bphi(\bx) \, , \quad x_1 > 0 .
\end{equation*}
Again, the solutions  of the corresponding Heisenberg equations 
are well known. The resulting field $\phi_g(t, \bx)$, where $t$ now
denotes the
proper time of the accelerated observer and $\bx = (x_1,x_2,x_3)$ the 
initial position of the field  (Rindler coordinates),   
is localized at the spacetime point 
$\bx_g(t) \doteq (\mbox{sh}(gt) x_1, \mbox{ch}(gt) x_1, x_2, x_3) \in
{\cal M}$ (inertial coordinates).
It can be represented in terms of the inertial field according to  
$\phi_g(t, \bx) = \phi_0(\bx_g(t))$. Thus for
any point in the wedge shaped region 
\mbox{${\cal W} = \{ x \in {\cal M} : x_1 > |x_0| \}$}  
(Rindler wedge) one has the equality 
$\phi_g(x) = \phi_0(x)$. Hence the physical interpretation of 
the field at a given spacetime point does not depend on 
the way how it was carried there, \ie on the particular observer. 

There exist many stationary states also for the accelerated observer. 
Of particular interest is the one--parameter family of 
quasifree states which, for given $\varsigma > 0$, are 
fixed by the two--point function, cf.\ \cite{HaNaSt}, 
\begin{equation*}
\w{\phi_g(t,\bx) \phi_g(s,\by)}_\varsigma \doteq
(2 \pi)^{-1} \! \int \! d \omega \, (1 - e^{ - \varsigma \, \omega})^{-1} 
e^{i \omega (s - t)} \, \int \! du \, C(\bx_g(0) - \by_g(u)) \,
e^{-i\omega u} \, ,
\end{equation*}
where $C$ is the commutator function given above. 
Taking averages
of the observables in ${\boldsymbol {\cal A}_{\, g}}$ 
with regard to the time evolution of the accelerated observer, 
their fluctuations disappear in all of these states 
in the limit of asymptotic times. Thus the states are  
not only stationary but also extremal. 

Since the states satisfy the Hadamard condition \cite{SaVe}  
and the Hadamard paramet\-rix coincides with that in
the inertial frame (restricted to ${\cal W}$), the 
accelerated observer has
at his disposal the same composite fields as the inertial
observer. In particular, he can make use of the 
thermometer observable introduced above. Its time 
evolution is given by 
 $\theta_g(t,\bx) = \theta_0(\bx_g(t))$,  hence 
$\theta_g(x) = \theta_0(x)$ for $x \in {\cal W}$, 
so its physical interpretation remains unchanged as well. The 
expectation values of the thermometer observable 
can be evaluated and are given by 
\begin{equation} \label{temperature}
\w{\theta_g(t,\bx)}_\varsigma = 
(12 x_1^2 g^2)^{-1} (\varsigma^{-2} - (g/2\pi)^2) \, .
\end{equation}

So the following picture emerges: For parameter values
$\varsigma < 2\pi / g$ the thermometer observable displays
the same value of the temperature all over any given hyperbolic hypersurface
$\{ y \in {\cal W} : y_1^2 - y_0^{2} = x_1^2 \}$. 
If $x_1 >0 $ is made smaller, \ie if 
the hypersurface is moved closer to the edge of ${\cal W}$ 
(the horizon of the observer) 
the temperature increases, whereas at large distances   
it tends to $0$. Thus the temperatures measured by the observer 
on his world line and on neighbouring orbits are 
different.\footnote{Compare the well known example of a 
self--gravitating star in equilibrium, where the 
local temperature also varies through 
the star \cite{To}. The thermometer observable $\theta_g$ registers
the analogous effect in the present context.} 
The validity of the zeroth law could in principle
be tested in this situation by an inertial observer whose
worldline is tangent to any given hyperbolic hypersurface 
and who prepares in his frame an equilibrium state of the 
corresponding temperature. If, at the meeting point, he  
takes over the thermometer from the accelerated observer 
and exposes it to his inertial equilibrium state
its readings will not change, according to theory. \ 
For parameter values $\varsigma > 2\pi / g$
the thermometer observable displays negative  
values. So there do not exist inertial equilibrium states  
which can coexist with the respective 
stationary states in the preceding sense.  
In other words, the latter states are only thermally stable  
in the presence of external accelerating forces. The special 
case $\varsigma = 2\pi/g$, corresponding to the Unruh scenario, will 
be discussed below.

All states $\w{\cdot}_\varsigma$
may be regarded as equilibrium states in the sense of the 
second law. To verify this one has to determine the energy transferred
to the states by cyclic operations in the accelerated
frame, cf.~relation (\ref{passivity}).
One finds (either by general arguments,  
making use of the built--in KMS--property of the states \cite{Ta,BrRo},
or by explicit computations) that the energy of these states 
always increases by such operations, \ie the 
states are passive. Thus the observer cannot take 
advantage of the temperature differences 
between the hyperbolic hypersurfaces in ${\cal W}$ in order to 
gain energy. These temperature differences, inducing 
corresponding pressure differences, serve to stabilize 
the states by compensating the 
accelerating forces which depend on the distance from the horizon. 

The fact that the states $\w{\cdot}_\varsigma$ are passive 
implies that they comply with another prominent 
property of equilibrium states: If one couples these
states locally to those of some small system 
(\eg an Unruh--DeWitt detector) by 
introducing a mild interaction term between the two separate   
dynamics, one can show that in the limit of large times  
the small system is driven to a state which is in equilibrium 
with the large system, cf.\ for example \cite{KoFrGiVe,BiMe}. More precisely,
the small system is passive in this limit and has the 
same Carnot--parameter as the large system before the coupling.
So one can use the small system in order to determine this
parameter. 

Without coupling the state  $\w{\cdot}_\varsigma$ 
to some external system, its Carnot parameter can 
also be determined as indicated in relation~(\ref{carnot}). 
The result is $C_\varsigma = \varsigma^{-1}$.  
This parameter determines the relation between the 
acceleration $x_1^{-1}$ felt by the observables at distance $x_1$
from the horizon and the temperature $\tau(x_1)$ 
which has to prevail there in order to attain global equilibrium.
Making use of relation (\ref{temperature}), 
one obtains in the present model  
$C_{\varsigma} = 
\mbox{\footnotesize $ g \, \sqrt{12 \, x_1^2 \, \tau(x_1) + (1/2
    \pi)^2}$}$. \ 
But it should be noticed that this relation is model dependent; 
it is different, for example, in massive free field theory.
Thus, whereas the Carnot parameter dictates the local conditions
for global equilibrium, it contains little information about the
local thermal properties of the states by itself. In particular, it 
may not be regarded as some kind of ``global temperature'' 
of the system in the presence of external forces. 

Having illustrated the general concepts entering into
the thermal interpretation of states in quantum field theory, let us 
finally discuss the Unruh effect. It presents itself for the 
special parameter $\varsigma = 2 \pi / g$. As 
was discovered in \cite{Fu,Se2},  
the corresponding functional $\w{\cdot}_{2 \pi / g}$ 
coincides with the Minkowskian vacuum  $\w{\cdot}_{\infty}$  on the algebra
${\boldsymbol {\cal A}_{\, g}}$.\footnote{For other values of $\varsigma$
there holds a similar, but somewhat weaker statement \cite{Ve}:
The restriction of any state $\w{\cdot}_\varsigma$ 
to the subalgebra of ${\boldsymbol {\cal A}_{\, g}}$,
generated by $\bphi$, $\bpi$ smeared with test functions 
in any given compact region of the 
half space $x_1 > 0$,  agrees with some state in 
Fock space. But the latter state depends on the size of this   
region and, with the exception of the vacuum, there is no
state in Fock space for which one has 
agreement on all of ${\boldsymbol {\cal A}_{\, g}}$.}
 In particular,   
\begin{equation*}
\w{\phi_g(t,\bx) \phi_g(s,\by)}_{2 \pi/g} =
\w{\phi_0(\bx_g(t) ) \phi_0(\by_g(s))}_\infty \, .
\end{equation*} 
The generally accepted interpretation of this equality 
is the assertion that a uniformly accelerated observer 
in the Minkowskian vacuum perceives
this state as an equilibrium state. 
But we can say more about its  specific thermal properties.

It follows from the preceding results that the empirical 
temperature of the vacuum in the accelerated frame is given by   
$ \w{\theta_g(t,\bx)}_{2 \pi/g} = 0 $ at all points in~${\cal W}$.
Thus it coincides with the temperature in the inertial frame,
contrary to the interpretation of the Unruh effect advocated in 
\cite{Un}. The indisputable and intriguing message of the latter article  
is the observation that the vacuum is passive also for the
accelerated observer. Despite the presence of external forces,  
the observer cannot extract energy from the state  
by cyclic operations. 
In a sense, this feature of the vacuum may be regarded as 
a kind of super--stability.

\addtolength{\textheight}{12mm}  
\addtolength{\voffset}{-12mm}    %
It remains to 
discuss the significance of the Carnot parameter in the accelerated frame. 
In an inertial frame, a Carnot process operating between 
temperatures $\tau_2 > \tau_1 \geq 0$ has the efficiency 
$\eta_{\, 0} = 1 - C_{\tau_1}/C_{\tau_2} = 1 - 
\mbox{\footnotesize  $\sqrt{ \tau_1 \, / \, \tau_2}$}$. 
Hence if one uses the vacuum as colder reservoir, $\tau_1 = 0$, one
could attain in principle the maximal possible efficiency~$1$. 
If one considers the analogous experiment in the accelerated frame
close to the world line of the observer, one gets 
for the corresponding efficiency 
$\eta_{\, g} = 1 - C_{\varsigma(\tau_1)}/ C_{\varsigma(\tau_2)} = 1 - 
\mbox{\footnotesize  $\sqrt{ (12 \tau_1 + (g/2\pi)^2) \, / \, 
(12 \tau_2+ (g/2\pi)^2)  }$}$. It  is strictly smaller than the
inertial one and decreases with increasing $g$. This result indicates 
that in the accelerated frame only such cyclic 
processes are possible for the given 
temperatures $\tau_1, \tau_2$ whose efficiency stays well below  
the ideal one of the Carnot process. In other words, due to the 
presence of external forces, optimal Carnot processes  
are impossible in the accelerated frame 
as a matter of principle. This reduction of efficiency affects also  
processes involving the vacuum as a thermal reservoir. 
It finds its expression in the increase of its Carnot parameter. 

Whereas in accelerated frames the relation between the local temperature 
and the Carnot parameter in general depends on the 
microscopic dynamics, this is not so for vacuum states.
Irrespective of the dynamics, their Carnot parameter is given
by the universal value $C_0 = g / 2 \pi$, depending only on
the acceleration $g$. Thinking of theories with 
several vacuum states it follows that an 
accelerated observer cannot use pairs of different vacua 
as thermal reservoirs for energy production. This 
fact is in conflict with the idea (reviewed in \cite{CrHiMa})
that, for an accelerated observer, the vacuum is filled with a
stream of ``particles'' carrying thermal energy. 
For this energy would not only depend on the external 
forces, causing the acceleration, but also on detailed
properties of the particles, such as their masses 
and their interaction. Hence the ability of 
the vacuum to coexist in accelerated frames with all
other vacuum states would be affected, in conflict with theory.
So the vacuum is as ``empty'' for accelerated 
observers as for inertial ones; its Carnot parameter 
merely indicates the external acceleration which 
acts on the observer and his equipment and narrows his
ability to perform cyclic processes. 

We emphasize that the preceding computations, performed 
in an appropriate algebraic framework, are consistent with  
those done in other settings of quantum field theory,
cf.~\cite{Ta,CrHiMa}. 
It is only the particular interpretation of the mathematical results,
reviewed in those articles,   
which is at odds with the physical picture emerging 
from the present discussion. 
The upshot of our analysis is the insight that,
within the present context, 
the thermal interpretation of states requires a more 
careful application of the basic theoretical 
concepts: The notions of equilibrium 
(passivity) and of efficiency of cyclic processes 
(entering in the Carnot parameter)
depend on the motion of the observer. This finds its 
formal expression in the appearance of the  
generator $\delta$ of the dynamics in the corresponding mathematical 
definitions. In contrast, the concept of empirical  temperature,
which is based on local kinematical observables 
allowing to compare different states, is independent of the 
state of motion of the observer. In the present analysis  
this found its expression in the equality 
$\theta_g = \theta_0$ of the accelerated and inertial 
``thermometers''. These facts have to be observed 
in a thorough interpretation of the Unruh effect and lead
to the present differing conclusions.

Our results suggest that, quite generally, 
the analysis of the thermal properties of states 
ought to be based on local (pointlike) observables 
in the presence of external forces 
since they maintain their inertial interpretation
for accelerated observers. 
This local point of view was advocated first in
\cite{BuOjRo} in a discussion of 
thermal properties of non--equilibrium states in Minkowski space.
In \cite{BuSch} the framework was extended in order to include
also curved backgrounds. The method was illustrated  
there on the example of observers in de Sitter space.
A systematic study and slight revision 
of this framework was presented  in \cite{So}
where it was also shown 
that the global Carnot parameter of passive states 
imposes in general relations between the acceleration,
the curvature and the temperature felt locally by an observer. 
A comprehensive list of references to 
other work, where this novel approach was adopted, may also 
be found in the latter article. 

\vspace*{10mm}
\noindent{\bf \Large Acknowledgments} \\[2mm]
We would like to thank Klaus Fredenhagen, Stefan Hollands, 
Hansj\"org Roos, Bert Schroer and Rainer Verch 
for constructive comments on a preliminary version of this article.

\end{document}